\documentclass[twocolumn,preprintnumbers,amsmath,amssymb,floatfix]{revtex4}
\usepackage{graphicx}
\begin{document}

\title{Controlling charge injection in organic field-effect transistors using self-assembled monolayers}

\author{B.\ H.\ Hamadani$^1$, D.\ A.\ Corley$^2$, J.\ W.\ Ciszek$^2$, J.\ M.\ Tour$^2$ and D.\ Natelson*$^{1,3}$}

\affiliation{$^1$ Department of Physics and Astronomy, Rice
University}

\affiliation{$^2$ Department of Chemistry and Smalley Institute
for Nanoscale Science and Technology, Rice University}

\affiliation{$^3$ Department of Electrical and Computer
Engineering, Rice University, 6100 Main St., Houston, TX 77005}

\begin{abstract}
We have studied charge injection across the metal/organic
semiconductor interface in bottom-contact poly(3-hexylthiophene)
(P3HT) field-effect transistors, with Au source and drain electrodes
modified by self-assembled monolayers (SAMs) prior to active polymer
deposition.  By using the SAM to engineer the effective Au work
function, we markedly affect the charge injection process.  We
systematically examine the contact resistivity and intrinsic channel
mobility, and show that chemically increasing the injecting electrode
work function significantly improves hole injection relative to
untreated Au electrodes.  
\end{abstract}

\maketitle

Improved understanding of the dynamics of charge motion at interfaces
between metals and organic semiconductors (OSCs) is crucial for
optimizing the performance of organic optoelectronic devices,
including organic field-effect transistors (OFETs).  In an OFET the
electronic structure of the OSC/contacting electrode interface can
strongly affect the overall performance of the device.  The band
alignment at the OSC/metal interface is influenced by several factors
such as interfacial dipole
formation\cite{Watkins2002,Knupfer2005,Ishii1999}, electrode
contamination\cite{Wan2005}, and OSC
doping\cite{Gao2001,Gao2003,PRB-Hamadani2005}.  Depending on the
particular band alignment, the charge injection mechanism can
significantly change, as seen by {\it e.g.} linear (ohmic) or
nonlinear current-voltage characteristics.

Ohmic contacts between metals and inorganic semiconductors are often
achieved by strong local doping of the contact regions, but such an
approach is challenging to implement in OSCs.  Bulk doping levels in
the OSC do affect injection.  In recent work\cite{PRB-Hamadani2005}
examining charge injection in bottom-contact OFETs based on
poly(3-hexylthiophene) (P3HT), we found that large contact resistances
and nonlinear transport at low dopant concentrations are consistent
with the formation of an increased injection barrier for holes.  Band
alignment is also significant\cite{Hamadani-JAP2005}.  For
example\cite{PRB-Hamadani2005}, the onset of nonohmic transport at low
doping is much more severe in devices with Au source and drain
electrodes than Pt.  It has previously been
shown\cite{Campbell1996,Nuesch1998,Zuppiroli1999,Tour2001} that, by
self-assembly of a layer of molecules with an intrinsic electric
dipole moment, the work function of metal electrodes can be lowered or
raised, affecting the size of the injection barrier at the metal/OSC
interface.  While limited attempts have been made to use this approach
to engineer contacts in OFETs\cite{Gundlach2001,Kim2003}, considerably
more effort has been dedicated to contacts in organic light emitting
diodes (OLEDs)\cite{Nuesch1998,Zuppiroli1999,deBoer2005} and
modification of the OSC/dielectric
interface\cite{Pernstich2004,Kline2006} in OFETs.  In contrast to
OLEDs, OFETs allow studies of transport with a single carrier type,
with carrier density tunable independently of the source-drain bias,
and with established procedures for discerning between contact and
bulk effects.  Minimizing contact resistances is particularly
challenging in planar OFETs since active contact areas are generally
much smaller than in vertical OLEDs.

In this letter, using channels of varying length, we systematically
examine the contact resistances and true channel mobility at various
doping levels of bottom-contact P3HT OFETs with Au electrodes modified
by self-assembly of dipolar molecular monolayers.  We correlate the
transport data with self-assembled monolayer (SAM) induced work
function changes as measured by scanning potentiometry.  In the case
of electron-poor (work function-raising) SAMs, we show that contact
resistances remain low compared to the channel resistance and the
transistors show linear transport.  These observations are consistent
with the ``pinning" of the local chemical potential at the interface
at an energy favorable to hole injection, and contrast sharply with
the strongly nonlinear injection observed at low doping levels in
OFETs made with bare Au electrodes.  Furthermore, devices with
electrodes modified by electron rich (work function-lowering) SAMs
show nonlinear transport and low currents at all doping levels,
becoming increasingly nonlinear as dopant density is reduced. This is
consistent with formation of an increased injection barrier for holes
in such devices.

OFETs are made in a bottom-contact configuration\cite{APL-Hamadani} on
a degenerately doped $p+$ silicon wafer used as a gate. The gate
dielectric is 200~nm of thermal SiO$_{2}$. Source and drain electrodes
are patterned using electron beam lithography in the form of an
interdigitated set of electrodes with systematic increase in the
distance between each pair. The channel width, $W$, is kept fixed for
all devices at $200~\mu$m. The electrodes are deposited by electron
beam evaporation of 2.5~nm of Ti and 25~nm of Au, followed by lift
off.  This thickness of metal is sufficient to guarantee film
continuity and good metallic conduction while attempting to minimize
disruptions of the surface topography that could adversely affect
polymer morphology.

\begin{figure}[!h]
\begin{center}
\includegraphics[clip, width=8cm]{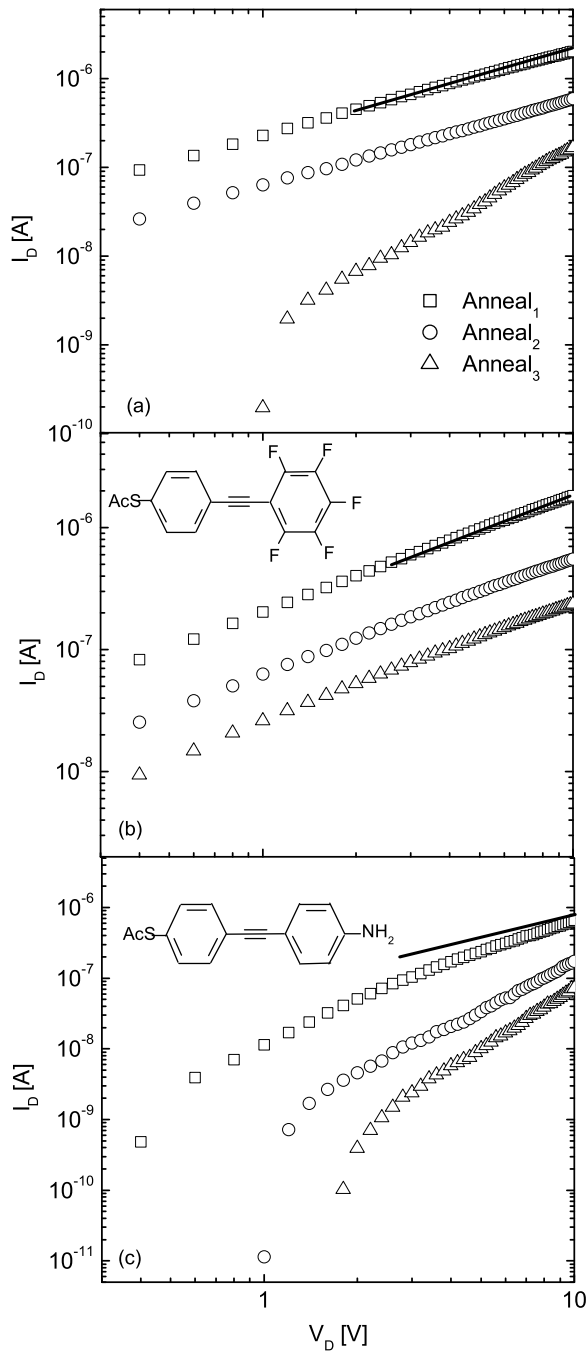}
\end{center}
\caption{(a) A log-log plot of the transport characteristics of a
Au/P3HT device for different annealing as described in the text.
(b) Similar plot for a Au/P3HT device with the electrodes modified
by F-OPE SAM shown in the inset. (c) Au/P3HT device with
electrodes modified by OPE-NH$_{2}$ SAM molecules shown in the
inset. For all devices, $L=40~\mu$m at $T=300$~K and at a fixed
$V_{G}=-70$~V with the same annealing schedule. The solid black
line has a slope of 1.} \label{fig1}
\end{figure}

Prior to SAM assembly, the substrates were cleaned for 2~min in an
oxygen plasma.  They were then immersed for about 24 h in a 1:1
ethanol-chloroform solution of the desired molecule at a
$\sim$0.25~mg/mL concentration, prepared under nitrogen gas. Three
types of molecules were used in this experiment: an electron poor
fluorinated oligo(phenylene ethynylene) (F-OPE) (see Fig. 1b inset),
and two electron-rich oligomers, OPE-NH$_{2}$ (Fig 1c inset) and
OPE-2(NH$_{2}$) (not shown, but similar to OPE-NH$_2$ with an
additional amine group immediately adjacent to the first).  These
molecules self-assemble from the thioacetate through standard Au-thiol
deprotection chemistry\cite{Cai2002}, and their characterization are
described in detail in the supplementary online material.  F-OPE
molecules are electron poor and upon assembly boost the metal work
function ({\it vide infra}), while the amine-terminated OPEs are
electron rich and are expected to have the opposite effect.

To characterize the effect of the SAM molecules on the effective
Au work function, we used a multi-mode atomic force microscope
(AFM) in the surface potential mode\cite{surf-pot} to measure the
surface potential difference between the SAM treated and bare Au
substrates. While not suited to absolute measurements of
work function, this method is useful for comparing relative
differences in work function between differently treated surfaces.
By comparing measured contact potentials of unmodified and
SAM-coated Au films, we found that the F-OPE treated Au substrates
exhibited an effective work function increased by $\sim$0.9~eV for
an assembly period of two days relative to untreated co-evaporated
Au films.  In addition, the F-OPE treated samples showed stability
and consistency in contact potential measurements over extended
periods (days) of exposure to ambient conditions.  For the
OPE-NH$_{2}$ and OPE-2(NH$_{2})$ treated surfaces, however, it was
difficult to obtain consistent surface potential differences with
respect to bare Au, though most showed a slight decrease
($\sim$0.1~eV) in work function.  These difficulties appear to
result from instability of the resulting surfaces under extended
exposure to ambient conditions.  However, as shown below, these
electron-rich molecules have a clear impact on band energetics at
the interface, with transport measurements suggesting the
formation of an injection barrier for holes.  

The organic semiconductor, 98\% regio-regular P3HT~\cite{Aldrich},
is dissolved in chloroform at a 0.06\% weight concentration,
passed through polytetrafluoroethylene (PTFE) 0.2\ $\mu$m filters
and solution cast onto the treated substrate, with the solvent
allowed to evaporate in ambient conditions. The resulting films
are tens of nm thick as determined by atomic force microscopy. The
measurements are performed in vacuum ($\sim$~$10^{-6}$ Torr) in a
variable temperature range probe station using a semiconductor
parameter analyzer. Exposure to air and humidity is known to
enhance hole doping in P3HT\cite{Hoshino2004}.  To reduce this
impurity doping, the sample is annealed at elevated temperatures
($\sim$350-380~K) in vacuum for several hours and then cooled to
room temperature for measurement. This results in a reduction in
the background dopant concentration as easily characterized
through the two-terminal bulk P3HT conductivity.

The devices operate as standard $p$-type FETs in accumulation
mode. With the source electrode grounded, the devices are measured
in the shallow channel regime ($V_{D}< V_{G}$).  Figure~1(a) shows
the transport characteristics of a Au/P3HT device with $L=40~\mu$m
at $T=300$~K and at a fixed $V_{G}=-70$~V for different doping
levels. In anneal$_{1}$, the sample was vacuum treated in the
analysis chamber at 300~K for 16 h. Anneal$_{2}$ corresponds
to the sample being further heated in vacuum for 18 h at
350~K, while anneal$_{3}$ includes yet an additional 18 h at
360~K.  As in earlier experiments\cite{PRB-Hamadani2005}, the
transport in this device with unmodified Au electrodes becomes
nonlinear at high annealing steps, and the current drops by orders
of magnitude.  We attributed this to the formation of an increased
injection barrier for holes, and similar effects have been
reported by others\cite{RepetAl03OE}.

In contrast, Fig. 1(b) shows the transport for a device with similar
geometric parameters and annealing schedule, in which the electrodes
were modified by F-OPE SAM molecules prior to P3HT deposition.  Even
though $I_{D}$ drops at higher annealing steps, the currents remain
linear with $V_{D}$ and as shown below, the contact resistance remains
much lower compared to bare Au devices.  This behavior is similar to
our previous observations\cite{PRB-Hamadani2005} for Pt/P3HT devices.
These effects have been verified in annealing cycles on multiple
arrays of F-OPE treated devices.

In Fig.~1(c), the electrode surfaces were modified by OPE-NH$_{2}$. In
this case, the currents are much lower than in either (a) or (b), and
even when the hole doping of the P3HT is significant, injection is
nonohmic, with $I_{D}$ rising super-linearly with $V_{D}$.  In highly
annealed conditions, this behavior is super-quadratic.  Transport data
for the OPE-2(NH$_{2}$) treated devices qualitatively looks very
similar to those in Fig.~1(c).

From the data in Figs. 1(a) and 1(b), we extracted the channel
resistance, $R_{ch}$, the intrinsic device mobility, $\mu$, and
the contact resistance $R_{c}$ from the L dependence of the total
device resistance, $R_{on}\equiv\partial V_{D}/\partial I_{D}$
over a $T$ and $V_{G}$ range as described in
Ref.\cite{APL-Hamadani}. We obtain $R_{c}$ in the limit $|V_{D}|<1
V$, where transport is still reasonably linear even after the
longer annealing runs.  We note that while we have developed a
procedure for extracting contact current-voltage characteristics
even in the limit of strong injection
nonlinearities\cite{Hamadani-JAP2005}, it is difficult to quantify
such injection by a single number such as $R_{c}$.  In the shallow
channel limit, it is straightforward to convert the gate
dependence of $R_{ch}$ into a field-effect
mobility\cite{APL-Hamadani}.

\begin{figure}[!h]
\begin{center}
\includegraphics[clip, width=8cm]{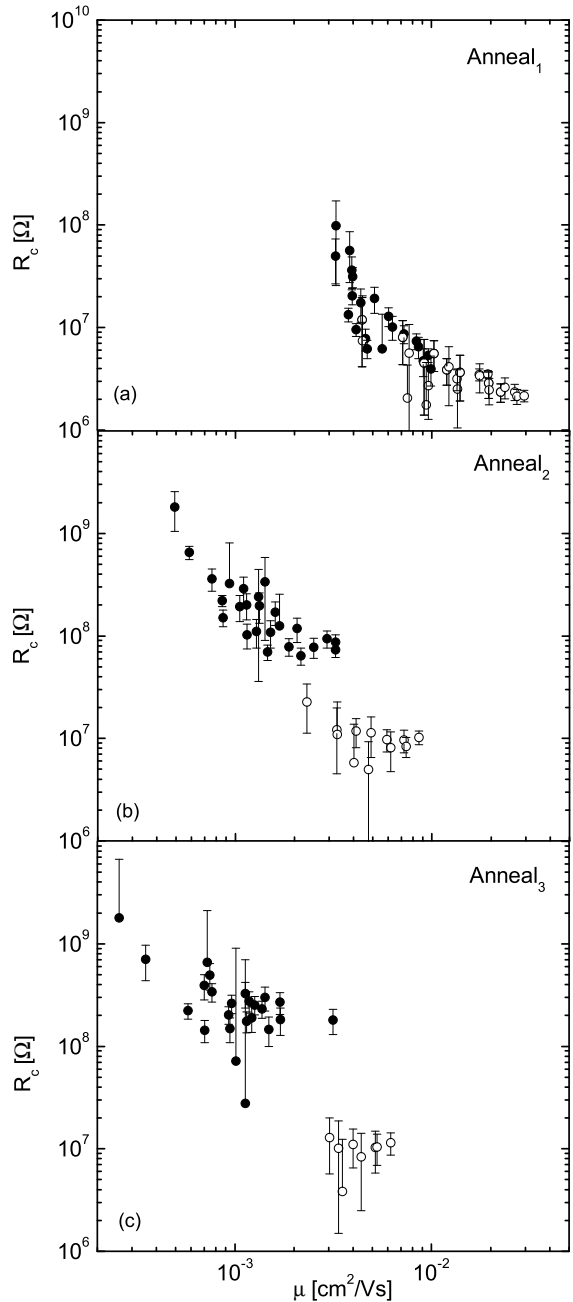}
\end{center}
\caption{(a) A log-log plot of $\mu$ vs $R_{c}$ for 2 sets of
devices over a series of temperatures and gate voltages. The open
symbols correspond to data from the F-OPE treated electrodes and
the filled symbols are extracted from bare Au/P3HT data.
Anneal$_{1}$ here corresponds to sample being pumped on in vacuum
at 320 K overnight (b) data retaken after sampled was annealed at
350 K for 18 h (anneal$_{2}$) (c) data taken again after
another anneal step similar to (b). The reason for fewer data points
in Fig. (b) and Fig. (c) for SAM treated sample is a smaller
contact resistance with significant error compared to the channel
resistance.} \label{fig2}
\end{figure}

Figure 2(a) shows a log-log plot of $\mu$ vs $R_{c}$ for two sets of
devices over a series of temperatures and gate voltages for an initial
annealing step. The open symbols correspond to data from the F-OPE
treated electrodes and the filled symbols are extracted from bare
Au/P3HT data. The error bars come from the uncertainty in the slope
and intercept of $R_{on}$ vs. $L$ plots\cite{APL-Hamadani}.  Indeed,
in device arrays with $R_{c}<<R_{ch}$, deviations from perfect scaling
of $R_{ch}\propto L$ can lead to ``best fit'' values of $R_{c}$ that
are actually negative (and hence cannot be plotted on such a figure),
though with appropriately large error bars.  Here, the mobility and
the contact resistance for both device sets are similar, consistent
with similarity in the magnitude of $I_{D}-V_{D}$ for both samples. In
the proceeding annealing steps, however, the contact resistance for
the sample with untreated electrodes increases significantly, compared
to the SAM treated device (Figs. 2(b) and (c)).  Whereas the Au/P3HT
devices become severely contact limited at high dedopings, the
treatment of Au electrodes with F-OPE molecules keeps the contact
resistance relatively low compared to channel resistance, and the
transport characteristics remain {\it linear}.  This ohmic injection
persists even when bulk $V_{G}=0$ conduction in the P3HT film is
completely suppressed at room temperature.

Our results in F-OPE treated devices are quantitatively similar to
those obtained in charge injection from Pt electrodes into
P3HT\cite{PRB-Hamadani2005}.  Although it is difficult to probe
the energy level alignment directly at the metal/organic interface
(due to the the thick P3HT film resulting from solution casting),
it is clear that increasing the Au effective work function results
in improved electronic performance of these OFETs. In light of the
many experiments showing the formation of interfacial dipoles at
the metal/OSC interface upon deposition of the
OSC\cite{Watkins2002,Knupfer2005,Ishii1999}, it is possible that
introduction of workfuction-raising SAMs such as F-OPE in our
experiments counteracts or prevents the work function-lowering
effect of these interfacial dipoles.  This can result in a
"pinning" of the energy levels at the interface such that there is
a small or non-existent injection barrier for holes.  On the other
hand, the work function-lowering OPE-NH$_{2}$ SAMs appear to
contribute more substantially to the interfacial dipoles,
resulting in significant Schottky barrier formation for holes at
the interface and severe nonlinear injection in these devices. The
subsequent dedopings have the same implications discussed
earlier\cite{PRB-Hamadani2005}.

One must also consider whether the different injection properties
could result from SAM-induced changes in the ordering of the P3HT at
the metal-OSC interface.  Such morphological differences may occur,
and would require careful interface-sensitive spectroscopies or
scattering measurements to confirm.  However, while improved P3HT
ordering at the F-OPE/P3HT interface would result in higher mobilities
and lower contact resistances, we find it unlikely that morphological
changes alone could explain the dramatic difference in injection
properties as a function of doping.  The data in Fig. 2 strongly
suggest significant differences in the band energetics between
the F-OPE treated and untreated Au electrodes.

We have used dipole-containing self-assembled monolayers on the Au
source and drain electrodes to strongly manipulate the charge
injection process across the metal-organic interface in a series of
polymer FETs based on P3HT.  To see the effect of dopant concentration
on device performance, we measure device properties after each of a
series of mild annealing steps in vacuum.  We extract the contact
resistances and the intrinsic channel mobility from the length
dependence of the resistance for bare Au/P3HT and flourinated-OPE
Au/P3HT devices where transport is still relatively linear at low
drain bias.  At low dopant concentrations, SAM-modified devices show
significantly lower contact resistances and higher mobilities compared
to unmodified devices. We attribute these findings to higher metal
work function and small injection barriers for holes in the case of
F-OPE SAM modified devices, resulting from better energetic alignment
with the valence band of the organic semiconductor.  These results
quantitatively demonstrate the power of simple surface chemistry in
modifying the dynamics of charge at interfaces with OSCs, even in
nearly undoped material.  Such techniques will be generally useful in
significantly improving technologies based on these versatile
materials.

D.N. acknowledges support from the David and Lucille Packard
Foundation, the Alfred P. Sloan Foundation, the Robert A. Welch
Foundation, and the Research Corporation.  J.M.T. acknowledges support
from DARPA and AFOSR.




\end{document}